\documentclass[preprint,amsmath,amssymb,aps,prd,showpacs]{revtex4-1}
\usepackage{epsfig}
\usepackage{dcolumn}
\usepackage{bm}
\usepackage{tikz}
\usepackage{graphicx, graphics, color}
\usepackage{dcolumn}
\usepackage{bm}
\usepackage{hyperref}
\usepackage[mathlines]{lineno}
\usepackage{float}
\usepackage{subfig}

\newcommand{\be}{\begin{equation}}
\newcommand{\ee}{\end{equation}}
\newcommand{\ben}{\begin{eqnarray}}
\newcommand{\een}{\end{eqnarray}}
\newcommand{\la}{{\lambda}}

\newcommand{\cD}{{\cal D}}

\newcommand{\cJ}{{\cal J}}
\newcommand{\cO}{{\cal O}}

\newcommand{\cE}{{\cal E}}
\newcommand{\cM}{{\cal M}}

\newcommand{\p}{\partial}
\newcommand{\na}{\nabla}

\newcommand{\tA}{\tilde A}
\newcommand{\tF}{\tilde F}

\newcommand{\ep}{\epsilon}

\newcommand{\ga}{\gamma}

\newcommand{\tB}{{\tilde B}}

\graphicspath{{figs/}}
\graphicspath{{../figs/}}

\newcommand{\vna}{\vec \na}


\begin{document} 

\title{Dark photon - dark energy stationary axisymmetric black holes}

\author{Marek Rogatko} 
\email{rogat@kft.umcs.lublin.pl}
\affiliation{Institute of Physics, 
Maria Curie-Sklodowska University, 
pl.~Marii Curie-Sklodowskiej 1,  20-031 Lublin,  Poland}


\date{\today}

\begin{abstract}
Using Ernst formalism, stationary axisymmetric black hole solution in Einstein-{\it dark matter}-{\it dark energy} gravity has been
elaborated. The {\it dark sector} was chosen as {\it dark photon} concept, where an auxiliary $U(1)$-gauge field coupled to ordinary Maxwell one was introduced,
while {\it dark energy} was modelled by the existence of positive cosmological constant.
Refining our studies to the case of vanishing cosmological constant,
the uniqueness theorem for the black hole in question has been proved.

 \end{abstract}

\maketitle
\flushbottom

\section{Introduction}
\label{sec:intro}
The nature of the illusive ingredient of our Universe mass, {\it dark sector}, is the most tantalising question in contemporary physics and astrophysics.
The influence of {\it dark matter} on galaxy rotation curves, motion of galaxy clusters, measurements of cosmic microwave background radiation,
baryonic oscillations \cite{ber18,ber18a}, as well as, pulsar timing array experiments \cite{sma23} has also been revealed.
The unknown {\it dark matter }sector constitutes almost
27 percent of the mass of the observable Universe \cite{planck}, while the most of the additional part of the mass comprises {\it dark energy}, responsible for the Universe expansion. The {\it visible} sector
constitutes only four percent of the Universe mass.

One of the simplest conceptual idea is to consider the existence of a {\it hidden sector} composed of particles weakly interacting with the ordinary matter.
The notion of {\it dark photon}, being hypothetical Abelian gauge boson coupled to the
ordinary Maxwell gauge field \cite{hol86}
 is very a plausible candidate for physics beyond the Standard Model.
The {\it dark photon} idea has been introduced in \cite{hol86} several years ago, however it acquires a contemporary justification in
the unification scheme  \cite{ach16}, where the mixing portals coupling Maxwell 
and auxiliary gauge fields are under intensive exploit, and the {\it hidden sector} states are charged under their own groups. 
On the other hand, it has been claimed that {\it dark photons} might be produced, e.g., during inflationary phase of the Universe evolution from
inflationary fluctuations \cite{gra16,sat22}, 
during reheating \cite{reheat}, from resonant creation during axion oscillations \cite{axionosc},
from dark Higgs \cite{dro19}, as well as, from cosmic strings \cite{cstrings}.

Furthermore, several anomalous astrophysical effects like $511~ keV$
gamma rays \cite{jea03}, excess of the positron cosmic ray flux in galaxies \cite{cha08}, and the observations of an anomalous monochromatic $3.56~ keV$ X-ray line in the spectrum of some galaxy clusters \cite{bub14},  may advocate the {\it dark photon} idea.
The other astrophysical observations and laboratory experiments \cite{fil20}, like for instance
studies of gamma rays emissions from dwarf galaxies \cite{ger15},
examination of dilaton-like coupling to photons caused by ultra-light {\it dark matter} \cite{bod15}, 
inspections of the fine structure constant oscillations \cite{til15}, {\it dark photon} emission taking pace during 1987A supernova event \cite{cha17}, 
 electron excitation measurements in CCD-like detector \cite{sensei}, the search for a {\it dark photon} in $e^+ e^-$ collisions at BABAR experiment \cite{lee14},
 measurements of the muon anomalous effect \cite{dav11},
 propose the possible range of values for {\it dark photon} - Maxwell field coupling constant and the mass of the hidden photon \cite{cap21}.
  
It turns out that {\it dark photon} acting as a portal to the {\it hidden sector}, which introduces {\it dark matter}
self-interactions, may constitute a solution of the small-scale structure problems \cite{spe00}. Moreover it can explain the XENONIT anomaly \cite{apr20}.
{\it Dark photons} can affect the primordial nucleon-synthesis altering the effective number of thermally excited neutrino degrees of freedom \cite{fra14},
and potentially influence on transport properties and exert
on stellar energy transport mechanism being the key factor during cooling neutron star processes \cite{lu22}.

It has been also reported that the new exclusion limit for the $\alpha$-coupling constant $\alpha =1.6 \times
10^{-9}$ and the mass range of {\it dark photon}
$2.1\times 10^{-7} - 5.7 \times10^{-6} eV$. These data were achieved by using the two state-of-art high-quality factor superconducting radio frequency cavities \cite{rom23}.
Improved limits on the coupling of ultralight bosonic {\it dark matter} to Maxwell photons, based on long-term measurements of two optical frequencies were
proposed in \cite{fil23}.

On the other hand, using quantum limited amplification,  the first probing of the kinetic mixing coupling constant to $10^{-12}$ level for majority of {\it dark photon} masses was given in \cite{ram23}, being the first stringent constraints on new {\it dark matter} parameters space.

Recently, it is to be noted that use of a cryogenic optical path and a fast spectrometer to study {\it dark photon} conversion into ordinary one at the metal surface plate, enables to
establish an upper bound on coupling constant  $\alpha < 0.3-2 \times 10^{-10}$ (at 95 percent confidence level) \cite{kot23}.

On the other hand, in the light of the first LIGO gravitational wave detection and Event Horizon Telescope (EHT) images of black hole shadow,
investigations of black holes and the influence of {\it dark sector} on their physics constitute
very interesting problem on its own.

The main aim of our paper is  to find the stationary axially symmetric solution to Einstein-Maxwell {\it dark photon} gravity with cosmological constant, sometimes 
being identified with {\it dark energy}. Additionally we treat the problem of the uniqueness of rotating black hole influenced by {\it dark matter} sector, where we restrict
our consideration to the case without {\it dark energy}.

Our paper is organized as follows. In Sec. II we consider the basic features of Einstein-Maxwell gravity replenished by the auxiliary $U(1)$-gauge field ({\it dark photon})
coupled to the Maxwell one by the so-called {\it kinetic mixing term}. Next, we derived the Ernst-like equations for stationary axisymmetric solution in 
the theory in question, adding to our inspection
the cosmological constant which will be identified with {\it dark energy}.
 In Sec. III, after introducing the adequate charges bounded with the gauge fields, one achieves the line element we are looking for.
 In Sec. IV we pay attention to the boundary conditions of the studied solution. Using the matrix type of Ernst equations one conducts the uniqueness proof for stationary axisymmetric black hole in the considered theory, when 
 cosmological constant is absent. Namely, we reveal that all
 stationary axisymmetric solutions to Einstein-Maxwell-{\it dark photon} gravity, being subject to the same boundary and regularity conditions, comprise
 the only black hole solution having regular event horizon with non-vanishing mass and $t,~\phi$ -components of $U(1)$-gauge fields.
  In the last section we concluded our investigations.

\section{Gravity with dark matter - dark energy sectors}
This section will be 
devoted to the basic features of Einstein-Maxwell gravity influenced by the {\it dark matter}  sector, which constitutes another $U(1)$-gauge field coupled
to the ordinary Maxwell one, by the so-called {\it kinetic mixing} term describing interactions of both gauge fields. Moreover we add the positive cosmological constant 
authorizing {\it dark energy} in the spacetime under inspection.
The action related to Einstein-Maxwell {\it dark photon} - {\it dark energy} gravity
 is provided by
 \be
S_{EM-dark~ photon-\Lambda} = \int \sqrt{-g}~ d^4x  \Big( R - 2\Lambda
- F_{\mu \nu} F^{\mu \nu} - B_{\mu \nu} B^{\mu \nu} - {\alpha}F_{\mu \nu} B^{\mu \nu}
\Big),
\label{act}
\ee  
where 
$\alpha$ is taken as a coupling constant between Maxwell and {\it dark matter} field strength tensors.

Introducing the redefined gauge fields $\tA_\mu$ and $\tB_\mu$, in the forms as follows:
\ben
\tA_\mu &=& \frac{\sqrt{2 -\alpha}}{2} \Big( A_\mu - B_\mu \Big),\\
\tB_\mu &=& \frac{\sqrt{2 + \alpha}}{2} \Big( A_\mu + B_\mu \Big),
\een
one can get rid of the {\it kinetic mixing } term obtaining the following:
\be
 F_{\mu \nu} F^{\mu \nu} +
B_{\mu \nu} B^{\mu \nu} +  \alpha F_{\mu \nu} B^{\mu \nu}
\Longrightarrow
 \tF_{\mu \nu} \tF^{\mu \nu} +
\tB_{\mu \nu} \tB^{\mu \nu},
\ee
where we have denoted $\tF_{\mu \nu} = 2 \p_{[\mu }\tA_{\nu ]}$ and respectively $\tB_{\mu \nu} = 2 \p_{[\mu }\tB_{\nu ]}$. 
Having it in mind,
the rewritten action (\ref{act}) implies
\be
S_{EM-dark~ photon-\Lambda}  = \int \sqrt{-g}~ d^4x  \Big( R - 2\Lambda
- \tF_{\mu \nu} \tF^{\mu \nu} - \tB_{\mu \nu} \tB^{\mu \nu}
\Big).
\label{vdc}
\ee
Variation of the action (\ref{vdc}) with respect to $g_{\mu\nu},~\tA_\mu$ and $\tB_\mu$ reveals the following equations of motion for Einstein-Maxwell {\it dark matter} 
gravity with the positive cosmological constant ({\it dark energy}):
\ben
G_{\mu \nu} + \Lambda g_{\mu \nu} &=&
2 \tF_{\mu \rho} \tF_{\nu}{}{}^{\rho } - \frac{1}{2}g_{\mu \nu}\tF_{\alpha \beta} \tF^{\alpha \beta}
+ 2 \tB_{\mu \rho} \tB_{\nu}{}{}^{\rho } - \frac{1}{2}g_{\mu \nu}\tB_{\alpha \beta} \tB^{\alpha \beta},\\
\na_{\mu} \tF^{\mu \nu } &=& 0, \qquad \na_{\mu} \tB^{\mu \nu } = 0.
\een
In what follows, one will focus on stationary axisymmetric solution in the considered theory of gravity. The corresponding 
line element is given by
\be
ds^2 = - a~e^{\frac{b}{2}}~(dt + \omega~d\phi)^2 + a~e^{-\frac{b}{2}}~d\phi^2 + \frac{e^{2u}}{\sqrt{a}}~(dr^2 + dz^2),
\label{metr}
\ee
where the functions appearing in the line element \eqref{metr} depend on $r$ and $z$-coordinates.

Further, we choose the following ansatze for $U(1)$-gauge fields, which will be also $r$ and $z$-dependent
\be
\tA= \tA_0~dt +\tA_\phi~d \phi, \qquad \tB = \tB_0~dt + \tB_\phi~d\phi.
\ee
For the brevity of the notation, we denote by 
$\vna k = (\p_r k,~\p_z k)$.

The Einstein-Maxwell {\it dark photon} equations of motion with cosmological constant $\Lambda$
modelled {\it dark energy}, are provided by
\ben \label{ein1}
\na^2 a &+& 2~\Lambda~e^{2u}~\sqrt{a} = 0,\\ \label{ein2}
4 a~\na^2 u &+& \na^2 a - e^b~a~(\vna \omega)^2 + \frac{1}{4}~a~(\vna b)^2 = 0,\\ \label{ein3}
- \frac{\vna \cdot (a~\vna b)}{2} &-& {a~e^b~(\vna \omega)^2} - {\omega \cdot \vna (e^b~a~\vna \omega)} \\ \nonumber
&=& {2}~\bigg[ e^{\frac{b}{2}} \bigg( (\vna \tA_0)^2~\omega^2 - (\vna \tA_\phi)^2 \bigg) - e^{- \frac{b}{2}}~(\vna \tA_0)^2 \bigg] \\ \nonumber
&+& {2}~\bigg[ e^{\frac{b}{2}} \bigg( (\vna \tB_0)^2~\omega^2 - (\vna \tB_\phi)^2 \bigg) - e^{- \frac{b}{2}}~(\vna \tB_0)^2 \bigg], 
\\ \label{ein4}
\vna \cdot (e^b~a~\vna \omega) &=& - 4 \omega~e^{\frac{b}{2}} ~\Big[ (\vna \tA_0)^2  + (\vna \tB_0)^2  \Big]
+ 4 e^{\frac{b}{2}} ~\Big(
\vna \tA_\phi \cdot \vna \tA_0 + \vna \tB_\phi \cdot \vna \tB_0 \Big).
\een
The differential operators $\vna$ and $\na^2$ appearing in the above set of equations,  are flat gradient and Laplacian operators written in $(r,~z)$-coordinates,
while the dots mean the scalar product of nabla operators.

On the other hand, the $t$ and $\phi$-components of Maxwell-{\it dark photon} relations imply respectively, for $\tA_\mu$ gauge field
\ben \label{m1}
\vna \cdot \bigg[ e^{-\frac{b}{2}}~\vna \tA_0 &+& \omega~e^{\frac{b}{2}}~\bigg( \vna \tA_\phi - \omega~\vna \tA_0 \bigg) \bigg] = 0,\\ \label{m2}
\vna \cdot \bigg[ e^{\frac{b}{2}}~\bigg( \vna \tA_\phi &-& \omega~\vna \tA_0 \bigg) \bigg] = 0,
\een
and for $\tB_\mu$ one
\ben \label{dm1}
\vna \cdot \bigg[ e^{-\frac{b}{2}}~\vna \tB_0 &+& \omega~e^{\frac{b}{2}}~\bigg( \vna \tB_\phi - \omega~\vna \tB_0 \bigg) \bigg] = 0,\\ \label{dm2}
\vna \cdot \bigg[ e^{\frac{b}{2}}~\bigg( \vna \tB_\phi &-& \omega~\vna \tB_0 \bigg) \bigg] = 0.
\een
Equations \eqref{m2} and \eqref{dm2} comprise the integrability conditions for a scalar potentials, say $A_3$ and  $B_3$. Namely they imply
\be
{\vec e}_\phi \times \vna A_3 = e^{\frac{b}{2}}~(\vna \tA_\phi - \omega~\vna \tA_0),
\label{bf}
\ee
and for $B_3$
\be
{\vec e}_\phi \times \vna B_3 = e^{\frac{b}{2}}~(\vna \tB_\phi - \omega~\vna \tB_0).
\label{dbf}
\ee
From the above relation we get the following relations for $A_3$ and $B_3$:
\be
\na_i \bigg(\ep^{i \phi m}~e_\phi~\na_m A_3 \bigg) = 0, \qquad \na_i \bigg(\ep^{i \phi m}~e_\phi~\na_m B_3 \bigg) = 0,
\label{eps}
\ee 
where we set $m = r,~z$,
the $\ep^{abc}$ in (\ref{eps}) denotes the Levi-Civita symbol in the orthonormal frame defined by the ordered triad $({\vec e}_{r},~{\vec e}_\phi,~{\vec e}_\theta)$.

It in turn leads to the conditions follows: 
$$\p_r~\p_z A_3 = \p_z~\p_r A_3, \qquad \p_r~\p_z B_3 = \p_z~\p_r B_3.$$
Extracting from the equation \eqref{bf}, the term ${\vec e}_\phi \times \vna \tA_\phi$, and calculating its diverges, we
obtain the following:
\be
\na_i \bigg( e^{-\frac{b}{2}}~\na^i A_3 - \omega~\ep^{i \phi m}~e_\phi~\na_m \tA_0 \bigg) = 0.
\ee
The same procedure applied to $B_3$ potential reveals
\be
\na_i \bigg( e^{-\frac{b}{2}}~\na^i B_3 - \omega~\ep^{i \phi m}~e_\phi~\na_m \tB_0 \bigg) = 0.
\ee

The above equations together with the relations \eqref{m1} and \eqref{dm1} will give us the set of Maxwell-{\it dark photon} -{\it dark energy} equations. Let us define 
the complex potential $\Phi_{(k)}$, where $k = \tF,~\tB$, in the form as
\be
\Phi_{(\tF)} = \tA_0 + i~A_3, \qquad \Phi_{(\tB)} = \tB_0 + i~B_3.
\label{potentials}
\ee
It can be easily seen that the Maxwell-{\it dark photon} equations may be rewritten by means of $\Phi_{(k)}$ as a single set of two complex equations
\be
\vna \cdot \bigg( e^{-\frac{b}{2}}~\vna \Phi_{(k)} - i~\omega~{\vec e}_\phi \times \vna \Phi_{(k)} \bigg) = 0.
\ee
On the other hand, we allow to apply the same procedure in respect to the equation \eqref{ein4}.
and cast it in the form as
\be
\vna \cdot \bigg[ e^b ~a~\vna \omega - 2 {\vec e}_\phi \times 
Im ( \sum_{k=\tF,\tB} \Phi_{(k)}^\ast~\vna \Phi_{(k)}) \bigg] = 0.
\label{ff}
\ee
Moreover, relation \eqref{ff} constitutes the integrability condition for the existence of the other potential, say $h$. Consequently, one achieves
\be
{\vec e}_\phi \times \vna h = e^b ~a~\vna \omega - 2 {\vec e}_\phi \times  
Im ( \sum_{k=\tF,\tB} \Phi_{(k)}^\ast~\vna \Phi_{(k)}),
\ee
while using relation (\ref{ff}), we obtain the relation given by
\be
{\vec e}_\phi \times \vna \omega = - \frac{1}{a e^b} \Big[ \vna h + 
2~Im~(\sum_{k=\tF,\tB} \Phi_{(k)}^\ast~\vna \Phi_{(k)}) \bigg) \bigg] = 0.
\ee
which in turn enables us to find that
the $(\phi,~t)$ component of Einstein-{\it dark photon} equations of motion may
be rewritten in terms of the potential $h$ and the complex potentials $\Phi_{(k)}$. Namely it implies
\be
\vna \cdot \bigg[ \frac{1}{a~e^{b}}~\bigg( \vna h + 
2~Im~(\sum_{k=\tF,\tB} \Phi_{(k)}^\ast~\vna \Phi_{(k)}) \bigg) \bigg] = 0.
\ee

Having in mind the above definitions, the relation \eqref{ein3} will be provided by
\ben
\frac{f}{a}~\vna \cdot (a~\vna f) &-& \vna f \cdot \vna f - f^2 \frac{\na^2 a}{a} = 2 f \sum_{k=\tF,\tB} \vna \Phi_{(k)} \cdot \vna \Phi_{(k)}^\ast \\ \nonumber
&-& 
\bigg[ \vna h + 
2~Im~(\sum_{k=\tF,\tB} \Phi_{(a)}^\ast~\vna \Phi_{(k)}) \bigg] \cdot \bigg[ \vna h + 
2~Im~(\sum_{k=\tF,\tB} \Phi_{(k)}^\ast~\vna \Phi_{(k)}) \bigg],
\een
where one denotes by $f = a~e^{b/2}$.

To proceed further, let us define the complex function given by the relation
\be
\cE = f -  \sum_{k=\tF,\tB} 
\Phi_{(k)}^\ast \Phi_{(k)} + i~h.
\ee
It happens that both Einstein-Maxwell {\it dark matter} - {\it dark energy} equations can be arranged in a system of the complex equations provided by
\ben \label{ern1}
\bigg( Re~\cE &+&  \sum_{k=\tF,\tB}
\Phi_{(k)}^\ast \Phi_{(k)} \bigg)~\frac{\vna  \cdot (a~\vna \cE)}{a} \\ \nonumber
&=&  
\bigg( \vna \cE + 2 \sum_{k=\tF,\tB}
\Phi_{(k)}^\ast  \vna \Phi_{(k)} \bigg) \cdot \vna \cE 
+ Re^2 \bigg( \cE + \sum_{k=\tF,\tB}
\Phi_{(k)}^\ast \Phi_{(k)} \bigg)~\frac{\na^2 a}{a}, \\ \label{ern2}
 \sum_{m=\tF,\tB} 
\bigg( Re~\cE &+&  \sum_{k=\tF,\tB}
\Phi_{(k)}^\ast \Phi_{(k)} \bigg)~\frac{\vna \cdot (a~\vna \Phi_{(m)})}{a}  \\ \nonumber
&=& 
\sum_{m=\tF,\tB} 
\bigg( \vna \cE + 2 \sum_{k=\tF,\tB}
\Phi_{(k)}^\ast  \vna \Phi_{(k)} \bigg) \cdot \vna \Phi_{(m)}.
\een
The above relations authorize the generalization of Ernst's equations describing Einstein-Maxwell system.
The real and imaginary parts of the first one, envisage the Einstein-Maxwell {\it dark matter} equations
with cosmological constant (sometimes interpreted as {\it dark energy}).
The real and imaginary parts
of \eqref{ern2} describe the Maxwell-{\it dark photon} equations of motion.
The aforementioned equations reduce to the ordinary complex Ernst differential relations for Einstein-Maxwell gravity, when one sets
the auxiliary gauge field equal to zero, as well as, assumes that the last term in (\ref{ern1}) vanishes. It yields that $a$ should be a harmonic function
$\na^2_{(r,z)} a=0$.

On the other hand, they can be achieved by varying the effective action $
S[\cE,\cE^\ast,\Phi_{(k)},\Phi_{(k)}^\ast]$, where one denotes by $k=\tF,\tB$. The aforementioned action yields
\ben \nonumber
S = \int dr ~dz \sum_{k=\tF,\tB} 
a&{}& \bigg[
\frac{ ( \na^i \cE + \Phi_{(k)} ^\ast \na^i \Phi_{(k)}  ) ( \na_i \cE + \Phi_{(k)} ^\ast \na_i \Phi_{(k)}  ) }{( \cE + \cE^\ast + \Phi_{(k)}  \Phi_{(k)} ^\ast )^2}
- \frac{\na_m \Phi_{(k)}  \na^m \Phi_{(k)} ^\ast}{( \cE + \cE^\ast + \Phi_{(k)}  \Phi_{(k)} ^\ast )} \\ 
&-&\frac{\na^j a}{2~a}~\frac{\na_j ( \cE + \cE^\ast + \Phi_{(k)}  \Phi_{(k)} ^\ast )}{( \cE + \cE^\ast + \Phi_{(k)}  \Phi_{(k)} ^\ast )}  \bigg].
\een

\section{Charging solution}
In Ref. \cite{ern68b} it was shown how to achieve the solution of the complex system of equations of the type given by relations \eqref{ern1}-\eqref{ern2}.
Namely, 
in order to find the form of the potentials $\Phi_{(k)} $ one should additionally assume that they are analytic functions and also analyze their asymptotic behavior.
Having all these in mind, by using the chain rule, we arrive at
\be
\frac{d^2 \cE}{d \Phi_{(\tF)}^2} \vna\Phi_{(\tF)} ~\Big( \vna\Phi_{(\tF)} \Big)^2 = \na^2 \cE~  \vna\Phi_{(\tF)} - \na^2 \Phi_{(\tF)}~\vna \cE,
\ee
and 
\be
\frac{d^2 \cE}{d \Phi_{(\tB)}^2} \vna\Phi_{(\tB)} ~\Big( \vna\Phi_{(\tB)} \Big)^2 = \na^2 \cE ~\vna\Phi_{(\tB)} - \na^2 \Phi_{(\tB)}~\vna \cE.
\ee
When one implements equation \eqref{ern1} multiplied and summed  by $\sum_{m=\tF,\tB} \vna\Phi_{(m)}$, and \eqref{ern2} multiplied by $\vna \cE$,
one obtains the relations for the potentials in the forms
\be
\forall_{\vna\Phi_{(\tF)} \neq 0}  \qquad \frac{d^2 \cE}{d \Phi_{(\tF)}^2} ~\Big( \vna\Phi_{(\tF)} \Big)^2 - Re~\Big(\cE +  \sum_{k=\tF,\tB} 
\Phi_{(k)}^\ast \Phi_{(k)} \Big) \frac{\na^2 a}{a}= 0,
\ee
and for $\Phi_{(\tB)} $ potential
\be
\forall_{\vna\Phi_{(\tB)} \neq 0}  \qquad \frac{d^2 \cE}{d \Phi_{(\tB)}^2} ~\Big( \vna\Phi_{(\tB)} \Big)^2 - Re~\Big(\cE +  \sum_{k=\tF,\tB} 
\Phi_{(k)}^\ast \Phi_{(k)} \Big) \frac{\na^2 a}{a} = 0.
\ee
In the next step
we decompose the complex Ernst potential $\cE$ in the form of a sum, where the first term does not contain $\Lambda$, while the second in $\Lambda$-dependent \cite{ast12}
\be
\cE = \cE_0 + \cE_{\Lambda}.
\ee
The forms of the equations reveal that there are no $\Lambda$-terms at zero order in in the cosmological constant, thus we get
\be
 \frac{d^2 \cE_0}{d \Phi_{(m)}^2} = 0,
\ee
which implies that to zero order in cosmological constant $\cE_0$ is linear function of the potential in question, i.e.,
\be
\cE_0 (\Phi_{(\tF)}) = f_0 + f_1~\Phi_{(\tF)}, \qquad \cE_0 (\Phi_{(\tB)}) = b_0 + b_1~\Phi_{(\tF)},
\ee
where $f_i, ~b_i$ are arbitrary constants. From the boundary conditions at infinity, i.e., $\Phi_{(m)} \rightarrow 0$ and $\cE_0 \rightarrow 1$,
we can fix $f_0$ and $b_0$ to be 1, while $
b_1 = -\frac{2}{Q_{(\tB)}}, ~ f_1 = -\frac{2}{Q_{(\tF)}}.
$
It implies the following forms of the potentials bounded with the visible and {\it dark} sectors
\be
\Phi_{(\tF)} = \frac{Q_{(\tF)}}{\xi + 1}, \qquad \Phi_{(\tB)} = \frac{Q_{(\tB)}}{\xi + 1},
\label{phi}
\ee
where we have set
\be
Q_{(\tF)} = \frac{\sqrt{2 - \alpha}}{2} \Big(Q_{(F)} - Q_{(B)}\Big), \qquad
Q_{(\tB)} = \frac{\sqrt{2 + \alpha}}{2} \Big(Q_{(F)} + Q_{(B)} \Big).
\ee
The quantities $Q_{(F)}$ and $Q_{(B)}$ are expressed in the standard way (as in charged Kerr metric derivation \cite{ern68b}) provided by
\be
Q_{(F)} = \frac{e_{(F)} + i g_{(F)}}{M}, \qquad
Q_{(B)} = \frac{e_{(B)} + i g_{(B)}}{M},
\label{char}
\ee
where  $e_{(i)}$ and $g_{(i)}$ are bounded respectively with electric and magnetic charges of Maxwell and {\it dark matter} sectors.
   
As was revealed in \cite{ern68a,ern68b,cha07,ast12}, the form  of $\cE$ for Kerr AdS/dS spacetime yields
\be
\cE = \frac{\xi -1}{\xi+1} + \frac{1}{\beta^2} \Big( (\xi + 1)^2 + q^2 \Big),
\ee
where $\xi = p x -iq y,~1/\beta^2 = \pm \Lambda M^2/3$ and respectively the other quantities are defined as \cite{ern68a}
\be
p = \frac{k}{M}, \qquad q = \frac{\hat a}{M}, \qquad x = \frac{r-M}{k}, \qquad y = \cos \theta, \qquad k = \sqrt{M^2 -{\hat a}^2}.
\ee
By $M$ we have denoted the total mass of the black hole, while $\hat a = J/M$ stands for its angular momentum per unit mass.

On the other hand, one arrives at the following expressions for the metric ingredients:
\ben
f &=& Re~\cE +  \sum_{k=\tF,\tB} \Phi_{(k)}^\ast \Phi_{(k)}
 \\  \nonumber
&=& \frac{ \xi^\ast \xi -1 + \mid Q_{(\tF)} \mid^2 + \mid Q_{(\tB)} \mid^2}{\mid \xi + 1 \mid^2}  +
\frac{1}{\beta^2} Re~\Big( (\xi +1)^2 + q^2 \Big)
\\ \nonumber
&=& \frac{\Delta_{(\tF,\tB)} - \Delta_\theta ~{\hat a}^2 \sin^2 \theta}{(r^2 + {\hat a}^2 \cos^2 \theta)}, \\
h &=& Im~ \cE = 2 \frac{ Im~\xi}{\mid \xi + 1 \mid^2}  + \frac{1}{\beta^2} Im~(\xi^2 + 2 \xi) \\
 &=& - 2 {\hat a} \cos \theta \Big( \frac{r}{\beta^2 M^2} + \frac{M}{r^2 + {\hat a}^2} \Big),
 \een
 while the other components of the rotating black hole with {\it visible} and {\it hidden sector} fields line element are provided by the following expressions:
 \ben
 \omega &=&\frac{{\hat a} \sin^2 \theta \Big(\Delta_{(\tF,\tB)} - \Delta_\theta (r^2 + {\hat a}^2) \Big)}
 {{\hat a}^2 \Delta_\theta \sin^2 \theta - \Delta_{(\tF,\tB)} } ,\\
 a &=& \sin \theta \sqrt{\Delta_\theta \Delta_{(\tF,\tB)} },\\
 e^{2u} &=& \sqrt{a} (r^2 + {\hat a}^2 \cos^2 \theta) ,\\
 e^{\frac{b}{2}} &=& \frac{\Delta_{(\tF,\tB)} - {\hat a}^2 \Delta_\theta \sin^2 \theta}
 {\sqrt{\Delta_\theta \Delta_{(\tF,\tB)}}(r^2 + {\hat a}^2 \cos^2 \theta) \sin \theta},
 \een
 where we have denoted, in the standard way, the quantities appearing in the above relations, i.e.,
 \ben
 \Delta_r &=& \Big( r^2 + {\hat a}^2 \Big)\Big( 1 - \frac{\Lambda}{3} r^2 \Big) - 2 Mr,\\
\Delta_\theta &=& 1 + \frac{\Lambda}{3} {\hat a}^2 \cos ^2 \theta,\\ \label{dark sect}
\Delta_{(\tF,\tB)} &=& \Delta_r + e_{(F)}^2 + g_{(F)}^2 + e_{(B)}^2 + g_{(B)}^2 + \alpha \Big( e_{(F)} g_{(B)} +   e_{(B)} g_{(F)} \Big).
\een 
 It can be seen
 that {\it dark matter} influences not only the line element by the squares of electric and magnetic charges , likewise the adequate Maxwell field ingredients,
 but also a mixture terms appear. They are connected with sum of electric Maxwell-magnetic {\it dark photon} and
 electric {\it dark photon}- magnetic Maxwell charges, with the proportionality constant $\alpha$ being the coupling constant between {\it visible} and {\it dark sectors}
 (see the action \eqref{act}).
 
 Moreover, the inspection of the equation (\ref{dark sect}) reveals that if we neglect {\it dark charge} $e_{(B)} =0$, then the main influence is exerted by {\it dark magnetic} charge
 $g_{(B)}$, and {\it dark magnetic} charge couples to the Maxwell electric one, i.e., $\alpha~ e_{(F)} g_{(B)}$.
 The same feature of the {\it dark sector} was also spotted in the analysis of the influence of {\it dark matter} on transport coefficients in chiral solids
 \cite{rog23a, rog23b}.

\section{Uniqueness theorem for dark matter stationary axisymmetric black hole}
In this section we pay attention to the problem of the uniqueness \cite{book} of stationary axisymmetric black holes with {\it dark sector}. On the other hand,
the uniqueness theorem for static axially symmetric black hole in magnetic Universe (say {\it dark Melvin Universe} ) 
was proved in Ref. \cite{rog23bh}.

In what follows
we shall restrict our considerations to the case when cosmological constant is equal to zero.
Namely, in relations  (\ref{ern1})-(\ref{ern2}), we set $\na^2_{(r,z)} a = 0 $ and $a=r$.
 
 It can be checked, by the direct calculations, that  defining homographic change of the variables, for the previously 
 defined quantities connected with both gauge fields, provided by
\be
\cE = \frac{ \xi -1}{\xi+ 1},
\ee
 and taking into account the relations (\ref{phi})-(\ref{char}), in the case of absence of the cosmological constant, the equations of motion (\ref{ern1}) and (\ref{ern2}), reduce
 to the single complex one of the following form:
 \be
 \Big( \xi^\ast \xi -1 + \mid Q_{(\tF)} \mid^2 + \mid Q_{(\tB)} \mid^2 \Big) \na^2 \xi
 = 2 \xi^\ast ~{\vec \na}\xi \cdot {\vec \na} \xi.
 \ee
 
 \subsection{Boundary conditions}
In order to study the relevant boundary conditions for stationary axisymmetric {\it dark photon} black hole one introduces the two-dimensional manifold $\cM$ \cite{car73},
with the spheroidal coordinates 
$r^2 = (\la^2 - c^2)~(1 - \mu^2)$, and $z = \la~\mu$,
where we set $\mu = \cos \theta$. In
the coordinates in question, the black hole event horizon boundary is situated 
at $\la = c$.
On the other hand, two rotation axis segments distinguishing the south and the north parts of the event horizon
are located at the limits $\mu = \pm 1$.
The obtained line element on two-dimensional manifold in question can be written in the form as follows:
\be
dr^2 + dz^2 = (\la^2 - \mu^2~c^2)\bigg(
{d\la^2 \over \la^2 - c^2} + {d\mu^2 \over 1 - \mu^2} \bigg).
\ee

Next, let us choose the domain of outer communication $<<\cD>>$ being a rectangle, which implies
\ben \label{dom}
\p \cD^{(1)} &=& \{ \mu = 1,~\la= c, \dots, R \},\\ \nonumber
\p \cD^{(2)} &=& \{ \la = c,~\mu = 1, \dots, -1 \},\\ \nonumber
\p \cD^{(3)} &=& \{ \mu = - 1,~\la= c, \dots, R \},\\ \nonumber
\p \cD^{(4)} &=& \{ \la  = R,~\mu= -1, \dots, 1 \}.
\een

As far as the boundary conditions \cite{car73,maz82} are concerned, at infinity, due to the asymptotic flatness of the solution we require that
$f/\la^2,~h,~\tA_0, ~A_3$ and $~\tB_0, ~B_3$ constitute well behaved functions of $1/\la$ and $\mu$, in the limit where $1/\la \rightarrow 0$ and the values of 
$\tA_0,~\tB_0$ tend respectively to $e_{(\tF)}$ and $e_{(\tB)}$, while the values of $A_3,B_3$ coincide with $g_{(\tF)}$ and $g_{(\tB)}$.
In terms of the above requirements they are given by
\ben \label{if1}
\tA_0 &=& - e_{(\tF)} \mu +  \cO(\la^{-1}), \qquad \tB_0 = - e_{(\tB)} \mu +  \cO(\la^{-1}), \\
A_3 &=& - g_{(\tF)} \mu +  \cO(\la^{-1}), \qquad B_3 = - g_{(\tB)} \mu +  \cO(\la^{-1}),\\
h &=& J \mu (3 -\mu^2) +  \cO(\la^{-1}),\\ \label{if4}
\frac{f}{\la^2} &=& (1 - \mu^2) (1 +  \cO(\la^{-1})).
\een

On the black hole event horizon,
where $\la \rightarrow c$, the quantities in question should behave regularly, i.e.,
they yield the following relations:
\ben \label{hor}
f &=& \cO(1), \qquad \frac{1}{f} = \cO(1),\\
\p_\mu A_3 &=& \cO(1), \qquad \p_\la A_3 = \cO(1), \\
\p_\mu \tA_0 &=& \cO(1), \qquad \p_\la \tA_0 = \cO(1),\\
\p_\mu B_3 &=& \cO(1), \qquad \p_\la B_3 = \cO(1), \\
\p_\mu \tB_0 &=& \cO(1), \qquad \p_\la \tB_0 = \cO(1),\\ 
\p_\mu h &=& \cO(1), \qquad \p_\la h = \cO(1).
\een
On the other hand, in the vicinity of the symmetry axis, where  $\mu \rightarrow 1$ (north polar segment) and $\mu \rightarrow -1$ (south polar segment),
one requires that $A_3, \tA_0,~B_3,~\tB_0, ~f,~h$ ought to be regular functions of $\la$ and $\mu$.
Consequently they are provided by
\ben \label{ax}
f &=& \cO(1 - \mu^2), \qquad \frac{1}{ f} \p_\mu f = 1+ \cO(1 - \mu^2),\\
\p_\la A_3 &=& \cO(1 - \mu^2), \qquad \p_\mu A_3 = \cO(1), \\
\p_\la \tA_0 &=& \cO(1- \mu^2), \qquad  \p_\mu \tA_0 = \cO(1), \\
\p_\la B_3 &=& \cO(1 - \mu^2), \qquad \p_\mu B_3 = \cO(1), \\
\p_\la \tB_0 &=& \cO(1-\mu^2), \qquad  \p_\mu \tB_0 = \cO(1), \\
\p_\mu h &+&
2 \Big( \tA_0 ~\p_\mu A_3 - A_3 ~\p_\mu \tA_0 \Big) + 2 \Big( \tB_0 ~\p_\mu B_3 - B_3 ~\p_\mu \tB_0 \Big) = \cO(1 - \mu^2),\\
\p_\la h &=& \cO((1-\mu^2)^2).
\een

\subsection{Uniqueness of Solutions}

It was revealed in Ref. \cite{gur82} that various combinations of the Ernst's equations of the type given by 
(\ref{ern1}) and (\ref{ern2}) can be comprised in matrix equation of the form as
\be
\p_{r} \Big[ P^{-1} \p_{r} P \Big]
+ \p_{z} \Big [ P^{-1} \p_{z} P\Big ] = 0,
\label{mat}
\ee
where by $P$ we have denoted $3\times 3$ Hermitian matrices with unit determinants.
Additionally it happens that for any constant, invertible matrix $A$, the matrix $A P A^{-1}$ is the solution of the relation (\ref{mat}),
enabling one to create all the transformations referred to the Ernst's system of partial differential equations.

Let us assume that the matrix $P$ components are enough differentiable in the domain of outer communication
$<<\cD>>$ of the two-dimensional manifold $\cM$, with boundary $\p \cD$. Suppose, further that we have
two different matrix solutions of the equation (\ref{mat}), $P_{1}$ and $P_{ 2}$, subject to the same boundary and differentiability
conditions, and consider
the difference between the aforementioned relations satisfies the equation of the form
\be
\na \Big(  P_{ 1}^{-1} \Big( \na Q \Big)
P_{ 2} \Big) = 0, 
\label{diff}
\ee
where one sets $Q = P_{1} P_{2}^{-1}$. 
Multiplying the equation (\ref{diff}) by
$Q^{\dagger}$ and taking the trace of the result, we achieve the following outcome:
\be
\na^2 q = Tr \Bigg[
\Big( \na Q^{\dagger} \Big) P_{ 1}^{-1}
\Big ( \na Q \Big) P_{ 2} \Bigg],
\label{tr}
\ee
where we set $q  = Tr Q $. Hermicity and positive definiteness of the matrix $P$,  allow us to postulate the matrix in the form as
 $P = M  ~M^{\dagger}$, which in turn 
 yields
 \be
\na^2 q = Tr \Big( \cJ^{\dagger} \cJ \Big),
\label{jj}
\ee
where $ \cJ= M_1^{-1} (\na Q ) M_{ 2}$.

Defining homographic change of the variables, for the previously defined quantities connected with both gauge fields, provided by
\be
\ep = \frac{ \xi -1}{\xi + 1}, \qquad \Psi_{(\tF)} = \frac{\eta_{(\tF)}}{\xi +1}, \qquad \Psi_{(\tB)} = \frac{\eta_{(\tB)}}{\xi +1}, 
\ee
enables us to find that the $P$ matrix implies
\be
P_{\alpha \beta } = \eta_{\alpha \beta} - \frac{2 \xi_\alpha~ {\bar \xi}_\beta}{<\xi_\delta~ {\bar \xi}^\delta>},
\ee
where we define the scalar product in the form as
\be
<\xi_\delta~ {\bar \xi}^\delta> = -1 + \sum \limits_\ga \xi_\ga~ {\bar \xi}^\ga, \qquad \ga= 1, \dots, q.
\ee
In the case under consideration $\xi^1 = \xi, ~\xi^2 = \eta_{(\tF)},~\xi^3 = \eta_{(\tB)}$ and $q=3$.

Moreover, for the brevity of the final notion, we change the notation in the relation (\ref{potentials}) for the following:
\be
\Phi_{(\tF)} = E_{(\tF)} + i~B_{(\tF)}, \qquad \Phi_{(\tB)} = E_{(\tB)} + i~B_{(\tB)}.
\label{potentials1}
\ee

The further step in the uniqueness proof of the {\it dark matter} stationary axisymmetric black hole solution is to find the trace $q = Tr(P_1 P^{-1}_2)$.
 Consequently, after some algebra, one arrives at
 \ben \label{qq} \nonumber
q =  P_{\alpha \beta (1)} P^{\alpha \beta}_{(2)} &=& 3 + \frac{1}{f_1 f_2} \Bigg \{
(f_1 - f_2)^2  + \Big[ 
\sum \limits_{i = \tF, \tB}  \Big( (E_{(i)1} -E_{(i)2})^2 + (B_{(i)1} - B_{(i)2})^2 \Big) \Big]^2 \\ \nonumber
&-& 2 (f_1 + f_2) \sum \limits_{i = \tF, \tB} \Big[ (E_{(i)1} -E_{(i)2})^2 + (B_{(i)1} - B_{(i)2})^2 \Big] \\ 
&+& \Big[ 2 \sum \limits_{i = \tF, \tB} \Big( B_{(i)1} E^{(i)}_2 - B_{(i)2} E^{(i)}_1 \Big) + (h_1 - h_2) \Big]^2 
\Bigg \}.
\een
Let us turn our attention to the relation (\ref{jj}) and integrate it over 
the domain of outer communication $<<\cD>>$ (\ref{dom}), using Stoke's theorem.
In accordance with the choice of $<<\cD>>$, one gets
\ben \label{bou} \nonumber
\int_{\p <<\cD>>} \na_k q~dS^k &=&   
\int_{\p <<\cD>>}~d\la~\sqrt{{h_{\la \la} \over h_{\mu \mu}}}~\p_\mu q \mid_{\mu = const}
+ \int_{\p <<\cD>>}~d\mu~\sqrt{{h_{\mu \mu} \over h_{\la \la}}}~\p_\la q \mid_{\la = const}
\\
&=&
\int_{\infty}^{c} ~d \la \sqrt{{h_{\la \la} \over h_{\mu \mu}}}~\p_\mu q \mid_{\mu = -1} + \int_{c}^{\infty} ~d \la \sqrt{{h_{\la \la} \over h_{\mu \mu}}}~\p_\mu q \mid_{\mu = 1} \\ \nonumber
&+&
\int_{1}^{-1} ~d \mu \sqrt{{h_{\mu \mu} \over h_{\la \la}}}~\p_\la q \mid_{\la = c} + \int_{-1}^{1} ~d \mu \sqrt{{h_{\mu \mu} \over h_{\la \la}}}~\p_\la q \mid_{\la \rightarrow \infty} \\ \nonumber
&=&
\int_{<<\cD>>} Tr \Big( \cJ^{\dagger} \cJ \Big)~dV.
\een
The behavior of the left-hand side of the above equation (\ref{bou}) will be elaborated by considering the integrals
over each
part of the domain of outer communication $ <<\cD^{(i)}>>$, where $i = 1,\dots,4$, chosen as a rectangle in the two-dimensional manifold with
coordinates $(\mu,~\la)$.

Namely, on the black hole event horizon, $\p \cD^{(2)}$, all the examined functions are well behaved, having asymptotic behavior given by $\cO(1)$. 
As $\la \rightarrow c$, the $r$-coordinate tends to $r \simeq \cO( \sqrt{\la -c})$ and the square root 
has the form of
$\sqrt{\frac{h_{\mu \mu}}{h_{\la \la}}} \simeq \cO( \sqrt{\la -c})$. 
Then, one can conclude that 
$\na_k q$ vanishes on the {\it dark matter} stationary axisymmetric black hole event horizon.

On the symmetry axis, $\p \cD^{(1)}$ and $\p \cD^{(3)}$,
when $\mu \pm 1$, all the quantities under inspection are of order $\cO(1)$. 
As $\mu \rightarrow 1$, $r$-coordinate tends to $\cO( \sqrt{1 - \mu})$, and for the case when $\mu \rightarrow - 1$, one has that 
$r \simeq \cO( \sqrt{1 + \mu})$. The behaviors of square roots are given by $\sqrt{\frac{h_{\la \la}}{h_{\mu \mu}}} \simeq \cO( \sqrt{1 + \mu})$, when $\mu \rightarrow -1$ and
$\sqrt{\frac{h_{\la \la}}{h_{\mu \mu}}} \simeq \cO( \sqrt{1 - \mu})$, for $\mu \rightarrow 1$. Thus the 
relations (\ref{bou})) reveals that $\na_m q = 0$, for $\mu \pm 1$.

For the case when $\la = R \rightarrow \infty$  all functions in question are well-behaved and have asymptotic behaviors given by the equations (\ref{if1})-(\ref{if4}).
On the other hand,  the square root in the considered limit tends to $\sqrt{\frac{h_{\mu \mu}}{h_{\la \la}}} \simeq \cO(\la)$. Inspection
of the boundary conditions given by the relations (\ref{if1})-(\ref{if4}) and (\ref{bou}), where we have differentiation with respect to $\la$,
reveal the fact that the studied integral tends to zero.

All the aforementioned arguments lead to the conclusion that
\be
\int_{<<\cD>>} Tr \bigg( \cJ_{(i)}^{\dagger} \cJ_{(i)} \bigg) = 0,
\label{jj1}
\ee    
which in turn implies that $P_{(i) 1} = P_{(i) 2}$, at all points belonging to the domain of outer communication, comprising a two-dimensional manifold
$\cM$ with coordinates $(r,~z)$.

It means that if one considers two stationary axisymmetric black hole solutions of Einstein-Maxwell {\it dark photon} gravity characterized respectively by
$(f_1,h_1,\tA_{0(1)}, A_{3 (1)}, \tB_{0 (1)}, B_{3 (1)})$ and 
$(f_2,h_2, \tA_{0 (2)}, A_{3 (2)}, \tB_{0 (2)}, B_{3 (2)})$, being subject to the same boundary and regularity conditions, they are identical.

In summary, the consequences of our research can be summarized as follows:\\
\noindent
{\bf Theorem}:\\
Consider a domain of outer communication $ <<\cD>>$ 
constituting a region of two-dimensional manifold with a boundary $ <<\p \cD>>$, equipped with the coordinate system $(r,~z)$ defined by 
$r^2 = (\la^2 - c^2)~(1 - \mu^2)$, and $z = \la~\mu$.

Assume further, that  $P_{(i)}$ are Hermitian positive, three-dimensional matrices, with unit determinants.
On the boundary of the domain $ <<\p \cD>>$, matrices $P_{(1)}$ and $ P_{(2) }$ authorize the solution of the equation
$$\p_{r} \Big[ P^{-1} \p_{r} P \Big]
+ \p_{z} \Big [ P^{-1} \p_{z} P\Big ] = 0,$$
and satisfy the relation $\na_m q =0$, where $q = Tr( P_1 P_2^{-1})$.

Then if $P_{(1)} = P_{(2)}$ in all domain of outer communication $ <<\cD>>$, implying that for at least one point $d \in <<\cD>>$,
 one arrives at the relation
 $ P_{(1) } (d) = P_{(2)} (d). $\\

Thus, all the stationary axisymmetric solutions of Einstein-Maxwell {\it dark photon} gravity subject to the same boundary and regularity conditions,
say Kerr-like {\it dark matter} black hole, comprise the only stationary axisymmetric symmetric black hole solution,
endowed with a regular event horizon, having non-vanishing
$\tA_{0},~A_{3 },~ \tB_{0},~B_{3}$ components of Maxwell {\it visible}  and {\it hidden sectors} gauge fields.
Having in mind equations introduced in Sec. II, the above components can be rewritten by means of $A_0,~A_\phi,~B_0,~B_\phi$ ones, i.e.,
\ben
\tA_0 &=& \frac{\sqrt{2 - \alpha}}{2} \Big( A_0 - B_0 \Big), \qquad \tB_0 = \frac{\sqrt{2 + \alpha}}{2} \Big( A_0 + B_0 \Big), \\
\tA_\phi &=& \frac{\sqrt{2 - \alpha}}{2} \Big( A_\phi - B_\phi \Big), \qquad \tB_\phi = \frac{\sqrt{2 + \alpha}}{2} \Big( A_\phi + B_\phi \Big).
\een

\section{Conclusions}
In our paper we have elaborated the stationary axisymmetric solution to Einstein-Maxwell {\it dark matter}-{\it dark energy} black hole solution.
The {\it dark sector} was modelled by {\it dark photon } theory, i.e., a new Abelian gauge field coupled to the ordinary Maxwell one, by means of
the so-called {\it kinetic mixing term}. On the other hand, the positive cosmological constant mandated phenomenologically the features of the influence of {\it dark energy}. The equations of
motion for the considered system were arranged into the form of Ernst-like system of complex relations.

The obtained metric components of the rotating Kerr-like {\it dark matter}-{\it dark energy} solution have envisaged the ordinary cosmological constant dependence
(like in Kerr dS spacetime \cite{cha07,ast12}), while the {\it dark sector} imprints its presence by square of electric Maxwell and {\it dark photon} charges, 
square of magnetic charges of both sectors and mixing electric-magnetic charges pertaining to {\it visible}/ {\it dark} and {\it dark}/{\it visible} sectors.

Then, we restrict our attention to the case of asymptotically flat solution and rearrange the adequate Ernst equations into the form of matrix equation.
Choosing the domain of outer communication $ <<\cD>>$ as a rectangle in two-dimensional manifold with coordinates $(r,~z)$, one shows that the two matrix 
solutions of the underlying equations being subject to the same boundary and regularity conditions are equal in $ <<\cD>>$.
Thus, one can draw the conclusion that Kerr-like {\it dark matter} stationary axisymmetric black hole solution to Einstein-Maxwell {\it dark photon}
gravity, authorizes the only stationary axisymmetric solution in the theory under inspection, having non-zero $A_0,~A_\phi,~B_0,~B_\phi$ gauge field components.

\acknowledgments
M. R. was partially supported by Grant No. 2022/45/B/ST2/00013 of the National Science Center, Poland.


\end{document}